\documentclass[12pt]{article}

\usepackage{amsmath}
\usepackage{amssymb}
\usepackage{latexsym}
\usepackage{graphics}
\usepackage{psfrag,fancyhdr,epsfig}

\begin{document}

\begin{titlepage}

\hfill hep-th/0609100\\

\vspace*{1.5cm}
\begin{center}
{\bf \Large Brane Cosmological Evolution with Bulk Matter }

\bigskip \bigskip \medskip

{\bf  C. Bogdanos and K. Tamvakis}

\bigskip

{ Physics Department, University of Ioannina\\
Ioannina GR451 10, Greece}

\bigskip \medskip
{\bf Abstract}
\end{center}
We consider the cosmological evolution of a brane for general bulk matter content. In our setup the bulk pressure 
and the energy exchange densities are comparable to the brane energy density. Adopting a phenomenological fluid ansatz and generalizations of it, we derive a set of
 exact solutions of the Friedmann equation that
exhibit accelerated expansion.
 We find that the effective equation of state
parameter for the dark energy can exhibit $w=-1$ crossing without the presence of exotic matter.

\end{titlepage}

\section{Introduction}

The present cosmological data\cite{R}\cite{WMAP} indicate that a significant percentage of the total energy of the universe is attributed to a 
component referred to as {\textit{``dark energy"}}. Dark energy is the driving force behind the observed accelerated expansion of the universe. Its physical origin 
is largely unknown to this point and the subject of intense theoretical speculation and research. Among the numerous cosmological models of dark energy are that of a
cosmological constant, exotic forms of matter which violate energy conditions, such as phantom fields\cite{PHANT} and quintessence\cite{QUINT}, as well as
modifications of gravitational theory itself\cite{DGP}. A cosmological constant is admittedly the simplest among these attempts, although the huge fine-tuning required for its
magnitude makes theorists very uneasy\cite{COSMO}. Independently of the challenge posed by the dark energy puzzle, in recent years, theories of extra spatial
dimensions, in which
the observed universe is realized as a {\textit{brane}} embedded in a higher dimensional spacetime ({\textit{bulk}}), have received a lot of attention. Ordinary
matter is confined on the brane but gravity is free to propagate in the entire spacetime\cite{A}\cite{AHDD}\cite{RS}. In these theories the cosmological evolution on the brane is described by an
effective Friedmann equation\cite{SMS} that incorporates non-trivially the effects of the bulk. Brane models open up new possibilities 
for the treatment of standing cosmological issues like the observed accelerated expansion\cite{BDL}.
 
The existence of a higher dimensional embedding space opens up the possibility of bulk matter. Bulk matter, since it certainly modifies the
 cosmological evolution on the brane, could possibly be a major contributor to the dark energy. Since, by assumption, we cannot probe the bulk, 
 the cosmological evolution on the brane, modified by the dark energy through its gravitational attributes, provides us also with a theoretical tool to study the
 bulk. It is also possible that the seemingly phantom-like characteristics of dark energy could arise as a result of the way that bulk dark matter affects
  the brane cosmological evolution, even though none of the matter constituents in either the bulk or the brane has these exotic features\cite{CGW}.  

It turns out that both the bulk pressure and the $05$ off-diagonal term of the bulk energy-momentum tensor can enter the cosmological equations and have a major
effect on the cosmological evolution on the brane. Models with energy exchange between the brane and the bulk have already been
 constructed\cite{KKTTZ} and this kind of interaction seems
capable of driving the acceleration under certain assumptions. In this article we consider the cosmological evolution of the brane in the presence of bulk matter. We
adopt a phenomenological description of bulk matter in terms of a fluid and derive exact solutions of the effective Friedmann equation with an accelerated expansion
profile. We find that, although the equation of state parameter of the bulk fluid is conventional, for suitable values of the bulk energy density and the dark
radiation constant, the effective equation of state parameter for the dark energy can cross the $``-1"$ line.

\section{General Framework}

Consider an Action of the general form
\begin{equation}
{\cal{S}} \,= \,\int {d^5 x\sqrt { - G} \left( {\,2M^3 R - \Lambda  +{\cal{L}}_B ^{(m)} } \right)\, + \,\int {d^4 x} \sqrt { -g} \left( { - \sigma  +
{\cal{L}}_{b} ^{(m)} } \right)} .
\end{equation}
$G_{MN}$ is the $5D$ metric with signature $(-,+,+,+,+)$ and $g_{\mu \nu}$ the 4D metric on the brane. $R$ is the five-dimensional
Ricci scalar, while $\Lambda$ is the bulk cosmological constant and $\sigma$ the positive brane tension.
We have included arbitrary matter content both in the bulk and on the brane through ${\cal{L}}_B ^{(m)}$ and ${\cal{L}}_{b} ^{(m)}$ 
respectively. M denotes the five-dimensional Planck mass. Varying the action with respect to the metric,
we obtain Einstein's equations
\begin{equation}
{\cal{G}}_{MN}\equiv\,R_{MN}  - \frac{1}
{2}G_{MN} R \,=\,\frac{1}{4M^3}\,T_{MN}\, {\label{Einstein1}}.
\end{equation}
The energy-momentum tensor $T_{MN}$ resulting from the above Action is of the form
\begin{equation}
T_{MN}\,=\,T^{( B )} _{MN} \, +\, T^{( b )} _{MN} \, - G_{MN} \Lambda  \,-g_{\mu\nu}\, \sigma \,\delta ( y
)\,\delta_M^{\mu}\delta_N^{\nu}\,{\label{e-m1}} ,
\end{equation}
where $T_{MN}^{(B)}$ results from ${\cal{L}}_B^{(m)}$ and $T_{MN}^{(b)}$ results from ${\cal{L}}_b^{(m)}$. 

Our metric ansatz will be of the form used in \cite{BDL}
\begin{equation}
ds^2\,=\,-n^2(y,t)dt^2\,+\,a^2(y,t)\gamma_{ij}dx^idx^j\,+\,b^2(y,t)dy^2
\end{equation}
or
\begin{equation}
G_{MN}\,=\,\left(\begin{array}{ccc}
-n^2(y,t)\,&\,0\,&\,0\\
0\,&\,a^2(y,t)\gamma_{ij}\,&\,0\\
0\,&\,0\,&\,b^2(y,t)
\end{array}\right)\,{\label{metric}}
\end{equation}
so that we have a Friedmann-Robertson-Walker brane geometry with a maximally symmetric 3-metric $\gamma_{ij}$. 
The fifth dimension is denoted by $y$ and the brane is situated at $y=0$. ${\cal{Z}}_{2}$ symmetry is assumed and $y$ takes values in the 
$[-\infty,\,+\infty]$ line. The metric functions $n(y,t)$, $a(y,t)$ and $b(y,t)$ are continuous
with respect to $y$ but may have discontinuous first derivatives at the location of the brane.

For a {\textit{bulk energy-momentum tensor}} $T_{MN}^{(B)}$ we shall take the matrix
\begin{equation}
{T^{(B)}}^{M}_{\,\,N}\,=\,\left(\begin{array}{ccc}
-\rho_B\,&\,0\,&\,P_5\\ 
\,0\,&\,P_B\delta^i_{\,\,j}\,&\,0\\
-\frac{n^2}{b^2}P_5\,&\,0\,&\,\overline{P}_B
\end{array}\right)\,\,,\,\,\,
T_{MN}^{(B)}\,=\,\left(\begin{array}{ccc}
\rho_Bn^2\,&\,0\,&\,-n^2P_5\\ 
\,0\,&\,P_Ba^2\gamma_{ij}\,&\,0\\
-n^2P_5\,&\,0\,&\,\overline{P}_Bb^2
\end{array}\right)\,{\label{bulk-e-m}} .
\end{equation}
Note the presence of the off-diagonal component $T^0_{\,\,5}=P_5$ signifying the flow of energy towards (or from) the brane. Note also the anisotropic
choice $\overline{P}_B\neq P_B$, in general.

For a {\textit{brane energy-momentum tensor}} $T_{MN}^{(b)}$ we shall take the matrix
\begin{equation}
{T^{(b)}}^{M}_{\,\,N}\,=\,\frac{\delta(y)}{b}\left(\begin{array}{ccc}
-\rho\,&\,0\,&\,0\\ 
\,0\,&\,p\delta^i_{\,\,j}\,&\,0\\
\,0\,&\,0\,&\,0
\end{array}\right)\,\,,\,\,\,
T_{MN}^{(b)}\,=\,\frac{\delta(y)}{b}\left(\begin{array}{ccc}
\rho n^2\,&\,0\,&\,0\\ 
\,0\,&\,pa^2\gamma_{ij}\,&\,0\\
\,0\,&\,0\,&\,0
\end{array}\right)\,.\,{\label{brane-e-m}}
\end{equation}

Substituting the above ansatze for the metric (\ref{metric}) and for the energy-momentum tensor (\ref{bulk-e-m}), (\ref{brane-e-m}) into the equations of
motion (\ref{Einstein1}), we obtain the equations

$$
 3\left\{\,{\frac{{\dot a}}
{a}\left( {\frac{{\dot a}}
{a} + \frac{{\dot b}}
{b}} \right) - \frac{{n^2 }}
{{b^2 }}\left( \,{\frac{{a''}}
{a}\, + \,\frac{{a'}}
{a}\left( {\frac{{a'}}
{a} - \frac{{b'}}
{b}} \right)}\, \right) \,+ \,k\frac{{n^2 }}
{{a^2 }}} \,\right\} $$
\begin{equation}\,= \frac{n^2}
{4M^{3}}\left\{\, \Lambda \, +\, \rho_B\,+\,\frac{\delta(y)}{b}\left(\,\sigma\,+\,\rho\,\right)\,\right\}\,,{\label{eom00}}
\end{equation}

$${\frac{{a^2 }}
{{b^2 }}\gamma _{ij} \left\{ {\frac{{a'}}
{a}\left( {\frac{{a'}}
{a} + 2\frac{{n'}}
{n}} \right) - \frac{{b'}}
{b}\left( {\frac{{n'}}
{n} + 2\frac{{a'}}
{a}} \right) + 2\frac{{a''}}
{a} + \frac{{n''}}
{n}} \right\}} 
$$

$$
 { + \frac{{a^2 }}
{{n^2 }}\gamma _{ij} \left\{ {\frac{{\dot a}}
{a}\left( { - \frac{{\dot a}}
{a} + 2\frac{{\dot n}}
{n}} \right) - 2\frac{{\ddot a}}
{a} + \frac{{\dot b}}
{b}\left( { - 2\frac{{\dot a}}
{a} + \frac{{\dot n}}
{n}} \right) - \frac{{\ddot b}}
{b}} \right\} - k\gamma _{ij} } 
$$

\begin{equation}
 =  \,\frac{{a^2 }}
{{4M^3 }}\left\{\,-\Lambda  + P_B\,+\,\frac{\delta(y)}{b}\left(\,p\,-\sigma\,\right)\,\right\}\gamma_{ij}\,,{\label{eomij}}
\end{equation}

\begin{equation}
3\left( {\frac{{n'}}
{n}\frac{{\dot a}}
{a} + \frac{{a'}}
{a}\frac{{\dot b}}
{b} - \frac{{\dot a'}}
{a}} \right) = - \frac{n^2}
{4M^{3}}\, P_5\,,
{\label{eom05}} 
\end{equation}

\begin{equation}
3\left\{ {\frac{{a'}}
{a}\left( {\frac{{a'}}
{a} + \frac{{n'}}
{n}} \right) - \frac{{b^2 }}
{{n^2 }}\left( {\frac{{\dot a}}
{a}\left( {\frac{{\dot a}}
{a} - \frac{{\dot n}}
{n}} \right) + \frac{{\ddot a}}
{a}} \right) - k\frac{{b^2 }}
{{a^2 }}} \right\} = \,\frac{{b^2 }}
{{4M^3 }}\left(\,-\Lambda \, + \,\overline{P}_B\,\right)\,, 
{\label{eom55}}
\end{equation}
where dots denote time derivatives and primes derivatives with respect to y. The curvature of the
maximally symmetric internal space is parametrized by $k=-1,0,1$.

At this point we may choose {\textit{Gauss normal coordinates}}, such that $b(t,y)=1$. In
addition, we have the freedom to take $n(0,t)=n_0(t)=1$. After theses simplifications the {\textit{junction conditions}} on the brane, 
obtained from (\ref{eom00}) and (\ref{eomij}), are
\begin{equation}
a'(+0,t)=-\frac{a_0}{24M^3}(\,\sigma\,+\,\rho\,)\,,\,\,\,\,n'(+0,t)=-\frac{1}{24M^3}(\,\sigma\,-2\rho\,-3p\,)\, ,{\label{jc3}}
\end{equation}
where $a_0(t)\equiv a(0,t)$ and $n_0(t)\equiv n(0,t)$. We have used ${\cal{Z}}_2$ symmetry taking $a'(-0,t)=-a(+0,t)$ and $n'(-0,t)=-n'(+0,t)$.  

Considering the last two equations of motion on the brane and using the junction conditions, we get
\begin{equation}
\dot{\rho}+3\left(\,\rho+p\,\right)\frac{\dot{a}_0}{a_0}=-2P_5\,,{\label{continuity}}
\end{equation}
\begin{equation}
\frac{\ddot{a}_0}{a_0}\,+\,\left(\frac{\dot{a}_0}{a_0}\right)^2\,+\,\frac{k}{a_0^2}\,=\,\frac{1}{(24M^3)^2}(\sigma+\rho)\left(2\sigma-\rho-3p\right)\,+\,
\frac{1}{12M^3}\left(\,\Lambda\,-\overline{P}_B\,\right)\,.{\label{friedmann1}}
\end{equation}
The first of these equations expresses energy conservation on the brane. For a non-vanishing $P_{5}$, energy exchange
between the brane and the bulk is present and the energy density $\rho$ on the brane is obviously not conserved. The
second equation will provide us with a modified Friedmann equation on the brane. Notice that the $``55"$ component of the bulk 
pressure appears on the right hand side of this equation (\ref{friedmann1}) affecting
the cosmological evolution.

\section{Friedmann Equation and Energy Conservation}

We proceed by considering equation (\ref{friedmann1}) which determines the time evolution of the scale factor on the brane. We shall assume an equation
of state $p=w\rho$ to hold between the energy density and pressure of matter on the brane. Dropping the subscript $``o"$ and 
defining $\beta \equiv (24M^3)^{-2}$ and $\gamma \equiv \sigma\beta$, we get \cite{KKTTZ}
\begin{equation}
\frac{\ddot{a}}{a}\,+\,\left(\frac{\dot{a}}{a}\right)^2\,+\,\frac{k}{a^2}\,=
\,\gamma\rho(1-3w)-\beta\rho^2(1+3w)-\frac{\overline{P}_B}{12M^3}+\frac{\lambda}{12M^3}\,,{\label{friedmann2}}
\end{equation}
where
\begin{equation}
\lambda\equiv \,\Lambda\,+\,\frac{\sigma^2}{24M^3}\,.
\end{equation}
We shall make the assumption that the effective $4D$ cosmological constant $\lambda$ vanishes when no matter is present. Thus, we shall proceed taking
$\lambda=0$. We shall also be interested in the {\textit{low density}} case, i.e.
\begin{equation}
(1+3w)\rho^2<<(1-3w)\rho\sigma\,.
\end{equation}

Our second order evolution equation (\ref{friedmann2}), with $\lambda=0$, can be cast in the form of two first order equations, 
one of which is analogous to the Friedmann equation of standard cosmology. This can be achieved by the introduction of a {\textit{dark energy variable}}
$\chi(t)$. Our new set of equations is
\begin{equation}
\left(\frac{\dot{a}}{a}\right)^2\,+\,\frac{k}{a^2}\,=
\,2\gamma\rho+\beta\rho^2\,+\chi\,-\frac{\overline{P}_B}{12M^3}\,,{\label{friedmann4}}
\end{equation}
\begin{equation}
\dot{\chi}+4\frac{\dot{a}}{a}\left(\chi\,-\frac{\overline{P}_B}{24M^3}\right)\,-\frac{\dot{\overline{P}}_B}{12M^3}\,-4P_5(\gamma
+\beta\rho)=0\,.{\label{friedmann5}}
\end{equation}
The pair of these equations\footnote{Comparing the first of these equations to our original evolution equation (\ref{friedmann2}), we have 
$$\frac{\ddot{a}}{a}=-\chi-(3w+2)\beta\rho^2-(3w+1)$$
Differentiating (\ref{friedmann4}) and using this equation proves (\ref{friedmann5}).} is equivalent (\ref{friedmann2}).

In addition to these two equations we also have the energy conservation equation (\ref{continuity}). Assuming the low density approximation, our set of
equations is
\begin{equation}
\dot \rho  + 3\rho \frac{{\dot a }}
{{a}}\left( {1 + w} \right) =  - 2P_5 ,
\label{conservation}
\end{equation}
\begin{equation}
\left(\frac{\dot{a}}{a}\right)^2\,+\,\frac{k}{a^2}\,=
\,2\gamma\rho\,+\,\chi\,-\frac{\overline{P}_B}{12M^3}\,,{\label{friedmannI}}
\end{equation}
\begin{equation}
\dot{\chi}+4\frac{\dot{a}}{a}\left(\chi\,-\frac{\overline{P}_B}{24M^3}\right)\,=\,\frac{\dot{\overline{P}}_B}{12M^3}\,+\,4P_5\gamma\,\,.{\label{friedmannII}}
\end{equation}

Equation (\ref{friedmannI}) is the {\textit{Friedmann equation}} describing cosmological evolution on the brane. 
The auxiliary field $\chi$ incorporates non-standard contributions ({\textit{dark energy}}) different than the regular matter content of the brane.
 The bulk matter contributes to the energy content of the brane through the bulk pressure terms ($\overline{P}_B=T^{(B)}_{55}$) that appear in the right
 hand side of the Friedmann equation. In addition to that, the bulk matter contributes to the energy conservation equation 
 (\ref{conservation}) through
 $P_5=T^0_5$ that accounts for energy moving from the brane to the bulk or the opposite. The functions $\overline{P}_{B}$, $P_5$ are functions of time 
 corresponding to the values of $\overline{P}_B(y,t)$ and $P_5(y,t)$ on the brane. The energy-momentum conservation
 $\nabla_MT^M_{\,\,N}=0$ cannot fully determine $\overline{P}_{B}$ and $P_5$ and a particular model of the bulk matter is required. In what follows
  we shall model the bulk matter in terms of a phenomenological regular bulk fluid ansatz that leads to analytic solutions. 
 \section{Digression on Bulk Matter}
 
 In this section we shall derive the conservation equations for the components of the bulk energy-momentum 
 tensor and investigate the constraints of the form of $P_5$
 and $\overline{P}_B$ on the brane for various ansatze and in the particular case of a fluid. Consider the conservation of the energy-momentum tensor
 \begin{equation}
 \nabla_MT^M_{\,\,N}=0\,\,\Longrightarrow\,\partial_MT^M_{\,\,N}\,+\,\Gamma_{MR}^MT^{R}_{\,\,N}\,-\Gamma_{MN}^RT^M_{\,\,R}\,=\,0\,.{\label{E-M-con}}
 \end{equation}
 The $``0"$ component of (\ref{E-M-con}) is
 \begin{equation}
\dot{\rho}_B+3\frac{\dot{a}}{a}(\rho_B+P_B)+3n^2P_5\left(\frac{n'}{n}+\frac{a'}{a}\right)+
n^2P_5'+\delta(y)\left\{\dot{\rho}+3\frac{\dot{a}}{a}(\rho+p)\,\right\}=0\,.{\label{E-M-con-0}}
\end{equation}
Integrating around $y=0$ and using the ${\cal{Z}}_2$ symmetry ($P_5(+0,t)=-P_5(-0,t)\equiv P_5(t)$), we reobtain the $``05"$ equation of motion on the
brane
$$\dot{\rho}+3\frac{\dot{a}}{a}(\rho+p)=-2P_5 ,$$
and, in the bulk, the equation
\begin{equation}
\dot{\rho}_B+3\frac{\dot{a}}{a}(\rho_B+P_B)+3n^2P_5\left(\frac{n'}{n}+\frac{a'}{a}\right)+
n^2P_5'\,=\,0\,.{\label{E-M-con-0-1}}
\end{equation}
Similarly, the $``5"$ component of (\ref{E-M-con}) gives
\begin{equation}
\overline{P}'_B\,+\,\dot{P}_5\,+\,P_5\left(\frac{\dot{n}}{n}+3\frac{\dot{a}}{a}\right)\,+\,\frac{n'}{n}\left(\overline{P}_B+\rho_B\right)\,=0\,.{\label{E-M-con-5}}
\end{equation}
In the limit $y\rightarrow +0$, these equations become
\begin{equation}
\dot{\rho}_B\,+\,3\frac{\dot{a}}{a}(\rho_B+P_B)\,+\,P'_5\,=\,\frac{1}{8M^3}(2\sigma-\rho-3p)P_5 ,{\label{con-1}}
\end{equation}
\begin{equation}
\dot{P}_5\,+\,3\frac{\dot{a}}{a}P_5\,+\,\overline{P}_B'\,=\,\frac{1}{24M^3}(\sigma-2\rho-3p)(\rho_B+\overline{P}_B).{\label{con-2}}
\end{equation}
These two equations as they stand cannot limit the freedom of choice for $\overline{P}_B(t)$ and $P_5(t)$, since the unknown functions $P'_5$ and
$\overline{P}_B'$ appear in them. Nevertheless, specific models for the bulk matter will greatly eliminate this freedom.

As a concrete model let us consider the energy-momentum tensor for a classical fluid in motion, moving at a speed $v$ along the fifth dimension
\begin{equation}
T^M_{\,\,N}=\,P_B^{(0)}\,\delta^M_{\,N}\,+\,(\rho_B^{(0)}+P_B^{(0)})\,U^MU_N\,,
\end{equation}
with
$$U^0=(1-v^2)^{-1/2},\,\,U^i=0,\,\,U^5=v(1-v^2)^{-1/2}.$$
Assuming a fluid-type equation of state
\begin{equation}
P_B^{(0)}=\omega\rho_B^{(0)}\,,
\end{equation}
we get
\begin{equation}
T^M_{\,\,N}\approx\rho_B^{(0)}\left(\begin{array}{ccc}
\omega-n^2(1+\omega)\,&\,0\,&\,v(1+\omega)\\
0\,&\,\omega\delta^i_{\,j}\,&\,0\\
-n^2v(1+\omega)\,&\,0\,&\,\omega+v(1+\omega)
\end{array}\right)\,\,+\,\,O(v^2)\,.
\end{equation}
In terms of the standard variables of the previous sections, we have
\begin{equation}
T^0_{\,\,0}=-\rho_B=-\rho_B^{(0)}\left[n^2(1+\omega)-\omega\right]\,,\,\,T^0_{\,\,5}=P_5=\rho_B^{(0)}v(1+\omega)\,,
\end{equation}
\begin{equation}
T^5_{\,\,0}=-n^2P_5=-n^2v(1+\omega)\rho_B^{(0)}\,,\,\,T^i_{\,\,j}=\delta^i_{\,\,j}P_B=\omega\rho_B^{(0)}\delta^i_{\,\,j}\,,
\end{equation}
\begin{equation}
T^5_{\,\,5}=\overline{P}_B=\rho_B^{(0)}\left[\omega+v(1+\omega)\right]\,.
\end{equation}
Substituting these quantities into equations (\ref{con-1}) and (\ref{con-2}) we obtain two equations involving $\dot{\rho}_B^{(0)}$ and ${\rho_B^{(0)}}'$.
Cancelling the spatial derivative among them, we obtain a single first order equation for $\rho_B^{(0)}(0,t)$, namely
$$\dot{\rho}_B^{(0)}\left[\omega+v(1+\omega)\right]\,+\,\rho_B^{(0)}\left\{3\frac{\dot{a}}{a}\omega(1+\omega)+3\frac{\dot{a}}{a}v(1+\omega)^2\right.$$
\begin{equation}\left.\,+\,\frac{v}{24M^3}\left[3\omega(1+\omega)(\rho+3p-2\sigma)\,+\,(1+\omega)^2 (\sigma-2\rho-3p)\,\right]\,\right\}\,=\,0\,.{\label{rho}}
\end{equation}
The lowest approximation solution to this is
\begin{equation}
\rho_B^{(0)}\,\approx\,\frac{C_B}{a^{3(1+\omega)}}\left(\,1\,+\,O(v\sigma/M^3)\,+\,O(v^2)\,\right)\,.{\label{aprox}}
\end{equation}
This corresponds to
\begin{equation}
\rho_B\,=\,\frac{C_B}{a^{3(1+\omega)}}\left(\,1\,+\,\dots\,\right)\,,\,\,\,P_B\,=\,\omega\frac{C_B}{a^{3(1+\omega)}}\left(\,1\,+\,\dots\,\right)\,,{\label{bulk1}}
\end{equation}
\begin{equation}
P_5\,=\,v(1+\omega)\frac{C_B}{a^{3(1+\omega)}}\left(\,1\,+\,O(v^2)\,\right)\,,\,\,\,\overline{P}_B\,=\,\omega\frac{C_B}{a^{3(1+\omega)}}\left(\,1\,+\,\dots\,\right)\,.{\label{bulk2}}
\end{equation}
The dots stand for $O(v\sigma/M^3)$ and $O(v^2)$.

From a phenomenological point of view it is possible to generalize this ansatz departing from its fluid interpretation. For instance, inserting
time-dependence in the parameter $v$ would introduce an extra $\dot{v}$ term in (\ref{rho}) but the lowest approximation (\ref{aprox}) would not be
affected.

\section{Exact Solutions and Late Time Acceleration}

Let us now return to our set of cosmological evolution equations in the low-density approximation $\beta\rho<<\gamma$
\begin{equation}
\left( {\frac{{\dot a}}
{a}} \right)^2 \, =\,\beta \rho ^2  + 2\gamma \rho  - \frac{k}
{{a^2 }} + \chi  - \frac{\overline{P}_B }
{{12M^{3}}}\,\approx\,2\gamma \rho  - \frac{k}
{{a^2 }} + \chi  - \frac{\overline{P}_B }
{{12M^{3}}}\,,{\label{coeq1}} 
\end{equation}
\begin{equation}
\dot \chi  + 4\frac{{\dot a}}
{a}\left( \chi  - \frac{\overline{P}_B }
{24M^{3}} \right)\,=\,4\beta \left( {\rho  + \frac{\gamma }
{\beta }} \right)P_5  + \frac{{\dot{\overline{ P}}_B }}
{{12M^{3}}}\,\approx\,4\gamma P_5   + \frac{{\dot {\overline{P}}_B }}
{{12M^{3}}}\,\,,{\label{coeq2}}
\end{equation}
\begin{equation}
\dot \rho  + 3\frac{{\dot a}}
{a}\rho \left( {1 + w} \right) \,=\,  - 2P_5\,\,\,\,\,\,.{\label{coeq3}} 
\end{equation}
Using the associated equation
\begin{equation}
\frac{{\ddot a}}
{a} =  - \left( {3w + 2} \right)\beta \rho ^2  - \left( {3w + 1} \right)\gamma \rho  - \chi\,\approx\, - \left( {3w + 1} \right)\gamma \rho  -
\chi\,,{\label{coeq4}}
\end{equation}
we can write a general expression for the {\textit{deceleration parameter}} $q$, which for $k=0$ is
\begin{equation}
q\equiv\,-\left(\frac{\dot{a}}{a}\right)^{-2}\left(\frac{\ddot{a}}{a}\right)\,=
\,\frac{\chi\,+\,(1+3w)\gamma\rho}{\chi\,+\,2\gamma\rho\,-\frac{\overline{P}_B}{12M^3}}\,\,.{\label{decce}}
\end{equation}
Accelerating behaviour emerges when $q<0$.

The time dependence of the scale factor $a(t)$ is determined by the Friedmann equation (\ref{coeq1}) which, ultimately, will have a general form
\begin{equation}
\left(\frac{\dot{a}}{a}\right)^2\,=\,C_0\,+\,\sum_{n=1}^NC_na^n\,+\,\sum_{n=1}^{\overline{N}}C_{-n}a^{-n}\,\,.{\label{genfried}}
\end{equation}
Assuming that we have an expanding behaviour, at late times the right hand side will be dominated by one term, namely
\begin{equation}
\left(\frac{\dot{a}}{a}\right)^2\,\approx\,C_{\nu}a^{\nu}\,\,.
\end{equation}
The resulting late time scale factor is either
$$a(t)\,=\,a(t_0)\left[\,1\,-\frac{\nu}{2}\sqrt{C_{\nu}}\left(a(t_0)\right)^{2/\nu}\,\left(t-t_0\right)\,\right]^{-\frac{2}{\nu}}\,,$$
for $\nu<0$, or
$$a(t)=a(t_0)\left(\frac{t_r-t}{t_r-t_0}\right)^{-\frac{2}{\nu}}\,,$$
for $\nu>0$. The time $t_r$ is the {\textit{``big rip"}} time for which the scale factor blows up. Accelerated expansion can be achieved in a trivial fashion when the dominant term is a
constant. In general, accelerated expansion requires $q=-1-\nu/2<0$ . It is interesting to see whether bulk matter can induce accelerated expansion in the case that 
no cosmological term is present ($\lambda=0$) and ordinary matter is in a standard $w=0$ or $w=1/3$ phase.

The set of cosmological equations (\ref{coeq1}), (\ref{coeq2}), (\ref{coeq3}) can be solved exactly for the following general ansatz for the bulk matter pressure on
the brane (see also \cite{CGW})
\begin{equation}
\overline{P}_B\,=\,D\,a^{\nu}\,\,,\,\,\,\,P_5\,=\,F\,\left(\frac{\dot{a}}{a}\right)\,a^{\mu}\,.{\label{Ansatz}}
\end{equation}
We shall proceed to obtain the resulting exact solution but we will justify this ansatz later. Substituting (\ref{Ansatz}) in equation (\ref{coeq2}), we obtain
$$\dot{\chi}+4\frac{\dot{a}}{a}\left(\chi-\delta a^{\nu}-\frac{{\nu}}{{2}}\delta a^{\nu}-\gamma F a^{\mu}\right)\,=\,0$$
or
\begin{equation}
\chi\,=\,\frac{{\cal{C}}}{a^4}\,+\,2\delta\frac{( \nu+2)}{(\nu+4)}a^{\nu}\,+\,\frac{4F\gamma}{(\mu+4)}a^{\mu}\,,{\label{eqxi}}
\end{equation}
where we have defined $\delta\equiv D/24M^3$. Similarly, we can proceed with the integration of (\ref{coeq3}) and get
\begin{equation}
\rho\,=\,\frac{\tilde{\cal{C}}}{a^{3(1+w)}}\,-\frac{2F}{\left[3(1+w)+\mu\right]}a^{\mu}\,.{\label{eqrho}}
\end{equation}
Substituting (\ref{eqxi}) and (\ref{eqrho}) into the Friedmann equation (\ref{coeq1}), we can write it in a conventional form as 
\begin{equation}
\left(\frac{\dot{a}}{a}\right)^2\,+\,\frac{k}{a^2}\,=\,\frac{8\pi}{3}G_N\,\rho_{eff}\,,{\label{Fried}}
\end{equation}
where $G_N=3\gamma/4\pi=3\sigma/4\pi(24M^3)^2$ is the $4D$ Newton's constant and the {\textit{effective energy density}} $\rho_{eff}$ stands for
\begin{equation}
\rho_{eff}\,=\,\frac{\tilde{\cal{C}}}{a^{3(1+w)}}\,+\,\frac{{\cal{C}}/2\gamma}{a^4}\,-\frac{2\delta}{\gamma(\nu+4)}a^{\nu}\,
+\,\frac{2(3w-1)F}{(\mu+4)\left[3(1+w)+\mu\right]}a^{\mu}\,.{\label{rhoeff}}
\end{equation}
Notice that for $w=1/3$ the $T^0_{\,\,5}$ decouples from $\rho_{eff}$.

The acceleration behaviour can be deduced from
\begin{equation}
\frac{\ddot{a}}{a}\,=\,-\frac{\gamma(1+3w)\tilde{\cal{C}}}{a^{3(1+w)}}\,-\frac{{\cal{C}}}{a^4}\,-2\delta\frac{(\nu+2)}{(\nu+4)}a^{\nu}\,
-\frac{2F\gamma(1-3w)(\mu+2)}{(\mu+4)\left[3(1+w)+\mu\right]}a^{\mu}\,.{\label{ddota}}
\end{equation}
For $\nu$ and $\mu$ greater than the standard matter exponent $-3(1+w)$, the bulk generated terms will dominate at late times. The corresponding late time 
deceleration
parameter comes out to be
\begin{equation}
q\,=\,\left\{\begin{array}{cc}
\nu>\mu\,&\,-1\,-\frac{\nu}{2}\\
\,&\,\\
\mu=\nu\,&\,-1\,-\frac{\nu}{2}\\
\,&\,\\
\mu>\nu\,&\,-1\,-\frac{\mu}{2}
\end{array}\right.\,\,.
\end{equation}
Note that the late time deceleration parameter comes out independent of the particular ratio of the ansatz parameters $D/F$. This means that
even if one of the two bulk pressure terms vanishes, the late time deceleration behaviour would be the same. This is not true of course for $\rho_{eff}$ which is 
proportional to $-4\delta+\frac{4\gamma(3w-1)F}{[3(1+w)+\nu]}$.

A justification for the ansatz (\ref{Ansatz}) that we employed can be found in the framework of the simple fluid model that we analyzed in the previous section. 
There, we found that an approximate solution for a conserved bulk energy-momentum tensor lead us to (\ref{bulk1}) and (\ref{bulk2}). The second of them reads
$$ P_5\,=\,v(1+\omega)\frac{C_B}{a^{3(1+\omega)}}\,,\,\,\,\overline{P}_B\,=
\,\omega\frac{C_B}{a^{3(1+\omega)}}\, . {\label{Ansatz1}}$$
As we mentioned there, we can adopt a phenomenological point of view and consider a velocity parameter that depends on time. A measure of temporal development on the brane is given by the Hubble parameter $H=\dot{a}/a$, so it would be natural to take $v=\zeta H$, where $\zeta$ is a
phenomenological parameter. With this choice, the above approximate solution takes the form
\begin{equation}
\overline{P}_B\,=\,\omega\frac{C_B}{a^{3(1+\omega)}}\,,\,\,\,\,P_5\,=\,\zeta(1+\omega)\frac{C_B}{a^{3(1+\omega)}}\left(\frac{\dot{a}}{a}\right)\,.{\label{Ansatz2}}
\end{equation}
Thus, for our phenomenological bulk fluid we obtain the effective energy density\footnote{Where $\overline{C}_B\equiv\frac{C_B}{24M^3}$ and
$\overline{\zeta}=24M^3\zeta$.}
\begin{equation}
\rho_{eff}\,=\,\frac{\tilde{\cal{C}}}{a^{3(1+w)}}\,+\,\frac{{\cal{C}}/2\gamma}{a^4}\,-\frac{2\omega}{\gamma(1-3\omega)}\frac{\overline{C}_B}{a^{3(1+\omega)}}\,
+\,\frac{2\overline{\zeta}(3w-1)(1+\omega)}{(1-3\omega)\left[3(w-\omega)\right]}\frac{\overline{C}_B}{a^{3(1+\omega)}}\,{\label{rhoeff1}}
\end{equation}
and
\begin{equation}
\frac{\ddot{a}}{a}\,=\,-\frac{\gamma(1+3w)\tilde{\cal{C}}}{a^{3(1+w)}}\,-\frac{{\cal{C}}}{a^4}\,+\,2\omega\frac{(1+3\omega)}{(1-3\omega)}\frac{\overline{C}_B}{a^{3(1+\omega)}}\,
+\,\frac{2\overline{\zeta}\gamma(1+\omega)(1-3w)(1+3\omega)}{(1-3\omega)\left[3(w-\omega)\right]}\frac{\overline{C}_B}{a^{3(1+\omega)}}\,.{\label{ddota1}}
\end{equation}
For $\omega<0$, the bulk term is dominant at late times and gives a decceleration parameter
\begin{equation}
q=\frac{1}{2}+3\frac{\omega}{2}\,.{\label{Acce1}}
\end{equation}
The corresponding acceleration range is
\begin{equation}
q<0\,\Longrightarrow\,-1\leq\,\omega\,<-\frac{1}{3}\,{\label{Acce2}}
\end{equation}

Both contributions to the Friedmann equation different than the brane energy-density, namely, the dark radiation term ${\cal{C}}/a^4$ and the bulk terms, can be 
interpreted as {\textit{dark energy}}. The effective equation of state parameter for the dark energy $w_{eff}^{(D)}$ can be computed from the prescription\cite{LJ}
\begin{equation}
w_{eff}^{(D)}\,=\,-1\,-\frac{1}{3}\frac{d\ln (\delta H^2)}{d\ln a}\,,{\label{eosp}}
\end{equation}
where $\delta H^2=H^2/H_0^2-\Omega_ma^{-3}$ accounts for all terms in the Friedmann equation not related to the brane matter, for which we have taken $w=0$. In our
case, equation (\ref{eosp}) becomes
\begin{equation}
w_{eff}^{(D)}\,=\,-1\,+\frac{1}{3}\frac{\left\{\,4{\cal{C}}\,+3\frac{(1+\omega)}{(1-3\omega)}\left[-4\omega\,+\,\frac{4}{3}\overline{\zeta}\gamma\frac{(1+\omega)}{\omega}\,\right]
\overline{C}_B(1+z)^{3\omega-1}\,\right\}}{\left\{\,{\cal{C}}\,+\,
\frac{1}{(1-3\omega)}\left[-4\omega\,+
\,\frac{4}{3}\overline{\zeta}\gamma\frac{(1+\omega)}{\omega}\,\right]\,\overline{C}_B(1+z)^{3\omega-1}\,\right\}}\,.
\end{equation}
The denominator is proportional to the effective dark energy density and should be positive. In order to achieve $w_{eff}^{(D)}<-1$, we need

\begin{equation}
4{\cal{C}}\,+3\frac{(1+\omega)}{(1-3\omega)}\left[-4\omega\,+\,\frac{4}{3}\overline{\zeta}\gamma\frac{(1+\omega)}{\omega}\,\right]
\overline{C}_B(1+z)^{3\omega-1}<0.
\label{condition}
\end{equation}
Taking into account that the denominator is positive, we obtain the following lower bound for the above expression:

\begin{equation}
4{\cal{C}}\,+3\frac{(1+\omega)}{(1-3\omega)}\left[-4\omega\,+\,\frac{4}{3}\overline{\zeta}\gamma\frac{(1+\omega)}{\omega}\,\right]
\overline{C}_B(1+z)^{3\omega-1}>(1-3\omega)\cal{C}.
\label{inequality}
\end{equation}
We want the lower bound to be negative. As already mentioned, in order to get accelerated expansion, $\omega$ must be within the range $-1\leq\omega<-\frac{1}{3}$. For this
range of values, we obtain $1-3\omega>0$. Thus, in order to get a negative lower bound, we must assume ${\cal{C}}<0$. Clearly, the existence of such a lower bound
is a necessary but not sufficient condition for $w=-1$ crossing to occur. However, with proper values for the parameters $\overline{\zeta}$ and $\overline{C}_B$ of the bulk
fluid the left hand side of (\ref{inequality}) can assume negative values as the redshift $z$ varies.

Summarizing, we presented a treatment of the cosmological evolution of braneworlds for general bulk matter content. In our framework the bulk matter pressure 
and the energy exchange between bulk and brane are comparable to the brane energy density. We adopted a phenomenological description of bulk matter in terms of a
fluid and using the ansatz ${T_B}^5_{\,\,5}\sim a^{-3(1+\omega)}$, ${T_B}^0_{\,\,5}\sim Ha^{-3(1+\omega)}$ we derived exact solutions of the Friedmann equation that
exhibit accelerated expansion. We found that the effective equation of state
parameter for the dark energy can exhibit $w=-1$ crossing without the use of matter that violates the energy conditions.

\bigskip

{\textbf{ Acknowledgments.}} We wish to thank Apostolos Dimitriadis for useful discussions. One of us (K.T.) thanks the CERN Theory Division for 
hospitality and Nicolas Tetradis for useful discussions. 
This research was co-funded by the European Union
in the framework of the Program $\Pi Y\Theta A\Gamma O PA\Sigma-II$ 
of the {\textit{``Operational Program for Education and Initial Vocational Training"}} ($E\Pi EAEK$) of the 3rd Community Support Framework
of the Hellenic Ministry of Education, funded by $25\% $ from 
national sources and by $75\%$ from the European Social Fund (ESF). C. B.
acknowledges also an {\textit{Onassis Foundation}} fellowship.


\begin{thebibliography}{99}

\bibitem{R} A. G. Riess et al., Astron. J. {\textbf{116}}, 1009 (1998); S. Perlmutter et al., Astrophys. J. {\textbf{517}}, 565 (1999);
 A. G. Riess et al.Astrophys.J. {\textbf{607}}, 665 (2004).
\bibitem{WMAP} D. N. Spergel, et al., WMAP Three Year Results: Implications for Cosmology, astro-ph/0603499.

\bibitem{PHANT} R. R. Calwell, Phys. Lett. B {\textbf{545}}, 23 (2002); R. R. Caldwell, M. Kamionkowski and N. N. Weinberg, Phys. Rev. Lett. {\textbf{91}}, 071301
(2003); J. M. Cline, S. Y. Jeon and G. D. Moore, Phys. Rev. D {\textbf{70}}, 043543 (2004).
 
\bibitem{QUINT} R. R. Caldwell, R. Dave and P. J. Steinhardt, Phys. Rev. Lett. {\textbf{80}}, 1582 (1998); P. J. E. Peebles and A. Vilenkin, Phys. Rev. D
{\textbf{59}}, 063505 (1999); P. J. Steinhardt, L. M. Wang and I. Zlatev, Phys. Rev. D {\textbf{59}}, 123504 (1999); M. Doran and J. Jaeckel, Phys. Rev. D
{\textbf{66}}, 043519 (2002); A. R. Liddle, P. Parson and J. D. Barrow, Phys. Rev. D {\textbf{50}}, 7222 (1994).

\bibitem{DGP} G. R. Dvali, G. Gabadadze and M. Porrati, Phys. Lett. B {\textbf{485}}, 208 (2000); G. R. Dvali, G. Gabadadze, M. Kolanovic and F. Nitti, Phys. Rev. D
{\textbf{64}} 084004 (2001).

\bibitem{COSMO} T. Padmanabhan, Phys. Rept. {\textbf{380}}, 235 (2003); V. Sahni and A. A. Starobinsky, Int. J. Mod. Phys. D {\textbf{9}}, 373 (2000); S. M. Carroll,
Living Rev. Rel. {\textbf{4}}, 1 (2001); S. Weinberg, Rev. Mod. Phys. {\textbf{61}}, 1 (1989).

\bibitem{A} I. Antoniadis, Phys. Lett. B {\textbf{246}}, 377 (1990).

\bibitem{AHDD} N. Arkani-Hamed, S. Dimopoulos and G. R. Dvali, Phys. Lett. B {\textbf{ 429}} (1998) 263; I. Antoniadis, N. Arkani-Hamed, S. Dimopoulos and G. R.
Dvali, Phys. Lett. B {\textbf{436}} 257 (1998).

\bibitem{RS} L. Randall and R. Sundrum, Phys. Rev. Lett. {\textbf{83}} (1999) 3370; Phys. Rev. Lett. {\textbf{83}} (1999) 4690.

\bibitem{SMS} T. Shiromizu, K. Maeda and M. Sasaki, Phys. Rev. D {\textbf{62}} (2000) 024012.


\bibitem{BDL} P. Binetruy, C. Deffayet and D. Langlois, Nucl. Phys. B {\textbf{565}} 269 (2000); P. Binetruy, C. Deffayet, 
U. Ellwanger and D. Langlois, Phys. Lett. B {\textbf{447}} 285 (2000).

\bibitem{CGW} R.G. Cai, Y. Gong and B. Wang, JCAP 0603 (2006); P. S. Apostolopoulos and N. Tetradis, hep-th/0604014.

\bibitem{KKTTZ} E. Kiritsis, G. Kofinas, N. Tetradis, T. N. Tomaras and V. Zarikas, JHEP {\textbf{0302}} (2003) 035; E. Kiritsis, N. Tetradis and T. N. Tomaras, 
JHEP {\textbf{0203}} (2002) 019; P. S. Apostolopoulos and N. Tetradis, Phys. Rev. D {\textbf{71}} 043506 (2005); P. S. Apostolopoulos and N. Tetradis, Phys. Lett. B
{\textbf{633}} 409 (2006); E. Kiritsis, JCAP {\textbf{0510}} 014 (2005); K. I. Umezu, K. Ichiki, T. Kajino, G. J. Mathews, R. Nakamura and M. Yahiro, Phys. Rev. D
{\textbf{73}} 063527 (2006).

\bibitem{LJ} E. V. Linder and A. Jenkins, Mon. Not. Roy. Astron. Soc. {\textbf{346}}, 573 (2003).


\end{thebibliography}
\end{document}